\documentclass[10pt,twocolumn,twoside]{IEEEtran} %double column
\usepackage{graphicx,cite,enumitem,}             		
			
\usepackage{amssymb,amsfonts,amsthm}   % Math packages

\usepackage[cmex10]{amsmath}
\usepackage{algorithm,caption}
\usepackage[noend]{algpseudocode}

\begin{document}
%
% paper title
% can use linebreaks \\ within to get better formatting as desired
% Do not put math or special symbols in the title.
\title{\huge Phase Noise Compensation Using Limited Reference Symbols in 3GPP LTE Downlink}
%
%
% author names and IEEE memberships
% note positions of commas and nonbreaking spaces ( ~ ) LaTeX will not break
% a structure at a ~ so this keeps an author's name from being broken across
% two lines.
% use \thanks{} to gain access to the first footnote area
% a separate \thanks must be used for each paragraph as LaTeX2e's \thanks
% was not built to handle multiple paragraphs
%

\author{Kun~Wang,~\IEEEmembership{Member,~IEEE,}
        Louay~M.A.~Jalloul,~\IEEEmembership{Senior Member,~IEEE,}
        and~Ahmad~Gomaa
        %and~Zhi~Ding,~\IEEEmembership{Fellow,~IEEE}% <-this % stops a space
\thanks{K. Wang is with Qualcomm Technologies, Inc., Santa Clara, CA. Email: kunwang@ieee.org.}% <-this % stops a space
\thanks{L. Jalloul and A. Gomaa are with Qualcomm Inc., San Jose, CA. Email: jalloul@ieee.org, aarg\_2010@yahoo.com.}% <-this % stops a space
%\thanks{Manuscript received April 19, 2005; revised December 27, 2012.}
}

\maketitle

% As a general rule, do not put math, special symbols or citations
% in the abstract or keywords.
\begin{abstract}
It is known that phase noise (PN) can cause link performance to degrade severely 
in orthogonal frequency division multiplexing (OFDM) systems, such as IEEE 802.11, 3GPP LTE and 5G. 
As opposed to prior PN mitigation schemes that assume perfect channel information 
and/or abundant reference symbols (e.g., in 802.11), the proposed PN compensation technique 
applies to LTE and 5G downlink with a limited number of reference symbols.
Specifically, in this work, we propose an efficient PN mitigation algorithm 
where the PN statistics are used to obtain more accurate channel estimates.  
Based on LTE downlink subframe structure, numerical results corroborate the effectiveness of the proposed algorithm.
\end{abstract}

% Note that keywords are not normally used for peerreview papers.

\vspace*{-2mm}

% For peer review papers, you can put extra information on the cover
% page as needed:
% \ifCLASSOPTIONpeerreview
% \begin{center} \bfseries EDICS Category: 3-BBND \end{center}
% \fi
%
% For peerreview papers, this IEEEtran command inserts a page break and
% creates the second title. It will be ignored for other modes.
%\IEEEpeerreviewmaketitle

% =========================================================
% =========================================================
\section{Introduction}
% 1. MIMO-OFDM and Phase Noise
\IEEEPARstart{M}{ultiple-input} multiple-output (MIMO) and orthogonal frequency division multiplexing (OFDM) are widely used in broadband data communications.
The capability of OFDM together with MIMO to achieve high-rate transmission by eliminating inter-symbol interference renders its adoption in several standards, including IEEE 802.11 Wireless Local Area Network (WLAN)? 3rd Generation Partnership Project (3GPP) Long-Term Evolution (LTE)  and 5G New Radio (NR).

% 2. Literature review
The phase noise (PN) introduced by the local oscillator degrades the high throughput performance of OFDM systems \cite{schenk2008rf}.
For this reason, extensive research efforts have been devoted to PN compensation in a variety of communication systems.
The works in \cite{wu2002phase,zou2007compensation,petrovic2007effects} use WLAN as the underlying system model that incorporates plentiful of reference symbols. The authors in \cite{zou2007compensation} presented a two-stage training-based algorithm, while the authors in \cite{petrovic2007effects} approximated the PN waveform using Fourier series. 
However, both \cite{wu2002phase} and \cite{petrovic2007effects} heavily rely on the impractical assumption of ideal channel state information (CSI) at receiver.
A few works \cite{corvaja2009joint,syrjala2009phase} on LTE-like systems have recently emerged in the literature as well. Nonetheless, the work \cite{corvaja2009joint} assumes pilot symbols in every OFDM symbol, which is not the case for real LTE systems, and the work \cite{syrjala2009phase} again assumes ideal CSI at the receiver.
In a recent work \cite{mehrpouyan2012joint}, joint estimation of channel and phase noise was proposed by using a decision-directed extended Kalman filter, but abundant training sequences and pilot symbols are needed in the system model presented. 
More recently, several works are put forward for PN compensation in mmWave systems  \cite{suyama2009iterative,zhang2015iterative}.

% 3. Our Contributions
Unlike previous works that either assume perfect CSI at receiver or rely on plentiful training/pilot signals,
our contribution in this work is a pragmatic scheme for PN compensation and data detection using only a limited number of reference signals which is applicable to the LTE and 5G downlink subframe structure.
%Instead of the rich block-type or comb-type pilots \cite{zou2007compensation}, the sparse reference symbols available in LTE systems pose a challenge.
Specifically, our novelty lies in the aspect that we view the PN contaminated channel as a whole and take the PN statistics into consideration in the 2-dimensional (2D) time-frequency minimum mean square error (MMSE) channel estimation, rather than trying to split PN from channel estimates as in the works \cite{zou2007compensation,corvaja2009joint}.
In particular, the proposed algorithm iteratively updates MIMO detection, phase noise estimation and channel estimation under a single objective.
Numerical results show the substantial gain in the link performance due to the iterative updates with the improvement in channel estimation by exploiting PN statistics.

\section{MIMO-OFDM System Model with Phase Noise} \label{sec:sys_model}
We consider a MIMO-OFDM communication system with $N_t$ transmit antennas, $N_r$ receive antennas and $N_c$ OFDM tones for transmission in a multi-path fading channel.
In this MIMO-OFDM system, carrier frequency offset is ignored due to the fact that PN is shown to be more dominant in the performance degradation \cite{corvaja2009joint,pollet1995ber}.
%Also, it is assumed that the cyclic prefix is longer than channel delay spread so that inter-block interference is completely eliminated.
Also, we restrict our attention to the PN at receiver only since transmitter PN can be approximated by an effective receiver PN \cite{schenk2005influence}.

Random PN generated by free-running oscillator is modeled as a Wiener process that exhibits Lorentzian-shape power spectrum density (PSD) \cite{schenk2008rf}. Let $\phi[n]$ denote the discrete Wiener process, then $\phi[n] = \phi[n-1] + \epsilon[n]$, where the increment $\epsilon[n]$ is Gaussian distributed with zero mean and variance $\sigma_{\epsilon}^2 = 4 \pi \beta T_s$. Here, $\beta$ is the single-sided $3$~dB bandwidth of the Lorentzian spectrum for the carrier process $e^{j \phi[n]}$ and $T_s$ is the sampling period. Note that, the received signals on all antennas undergo the same phase noise process, since a common local oscillator is used for all receive chains in the direct down-conversion to base band.
For the sake of notational simplicity, define $\mathcal{N}_t \triangleq \{ 1, 2, \ldots, N_t \},  \mathcal{N}_r \triangleq \{ 1, 2, \ldots, N_r \}$ and $\mathcal{N}_c \triangleq \{ 0, 1, \ldots, N_c-1 \}$.
Then, the PN contaminated frequency-domain signal at the $j$-th receive antenna on the $k$-th tone can be written as
\begin{equation}
y_j [k] = a[k] \circledast \sum_{i = 1}^{N_t} ( H_{ji}[k] \cdot x_i [k] ) + w_j[k], \quad k \in \mathcal{N}_c,
%j \in \mathcal{N}_r,
\end{equation}
where $\circledast$ denotes circular convolution, $x_i [k]$ is the transmit signal at antenna $i$ on the $k$-th tone and $w_j[k] \sim \mathcal{CN}(0,\sigma_w^2)$ is the additive white Gaussian noise at $j$th receive antenna.
Moreover, $a[k]$ is the frequency-domain phase noise, given by
\begin{equation}
a[k] = \frac{1}{N_c} \sum_{n=0}^{N_c-1} e^{j \phi[n]} e^{-j \frac{2 \pi k n}{N_c}}, \quad k \in \mathcal{N}_c,
\end{equation}
and $H_{ji}[k]$, the frequency-domain channel between transmitter $i$ and receiver $j$, is the DFT of time-domain channel impulse response $\{ h_{ji}[n] \}_{n=0}^{L-1}$, that is
\begin{equation}
H_{ji}[k] = \sum_{n=0}^{L-1} h_{ji}[n] e^{-j \frac{2 \pi k n}{N_c}}, \quad k \in \mathcal{N}_c.
\end{equation}

Then, the overall MIMO-OFDM transmission with phase noise contamination can be expressed as
\begin{equation} \label{eq:rx_sig1}
\mathbf{y} = (\mathbf{A} \otimes \mathbf{I}_{N_r}) \mathbf{H} \mathbf{x} + \mathbf{w},
\end{equation}
where $\otimes$ represents the Kronecker product and $\mathbf{I}_{N_r}$ is the $N_r$-by-$N_r$ identity matrix. In Eq.~(\ref{eq:rx_sig1}), the received signal $\mathbf{y} = [ \mathbf{y}[0]^T \, \mathbf{y}[1]^T \, \ldots  \mathbf{y}[k]^T \ldots \, \mathbf{y}[N_c-1]^T ]^T$, in which the signal vector $\mathbf{y}[k] = [y_1[k] \, y_2[k] \, \ldots \, y_{N_r}[k] ]^T$. The vectors $\mathbf{x}$ and $\mathbf{w}$ are defined similarly as $\mathbf{y}$. The circulant matrix $\mathbf{A} = \text{cir}(\mathbf{a})$ is the cyclic shift of the PN vector $\mathbf{a} = [ a[0] \, a[1] \, \ldots \, a[N_c-1] ]^T$. The channel matrix $\mathbf{H} = \text{blkdiag}
\{\mathbf{H}[k]\}_{k \in \mathcal{N}_c}$ with $\mathbf{H}[k] = [ H_{ji}[k] ]_{i \in \mathcal{N}_t, j \in \mathcal{N}_r}$ and $\text{blkdiag}(\cdot)$ represents block diagonal.
For ease of subsequent derivations, Eq.~(\ref{eq:rx_sig1}) is split for each tone, shown as follows
\begin{equation} \label{eq:rx_sig2}
\mathbf{y}[k] = a[0] \mathbf{H}[k] \mathbf{x}[k] + \sum_{r=0,r \neq k}^{N_c-1} a[k-r] \mathbf{H}[r] \mathbf{x}[r] + \mathbf{w}[k],
%\quad k \in \mathcal{N}_c,
\end{equation}
in which the first term $a[0]$ is referred to as common phase error (CPE), and the second term $\sum_{r=0,r \neq k}^{N_c-1} a[k-r] \mathbf{H}[r] \mathbf{x}[r]$ is called inter-carrier interference (ICI), 
each element of which is zero-mean with variance $\sigma_{ICI}^2 = 2 \pi \beta T_s N_t / 3$ for large $N_c$ \cite{wu2002phase,petrovic2007effects}.

% =========================================================
% =========================================================
\section{2D MMSE Channel Estimation with CPE Term} \label{sec:2d_mmse}
\begin{figure}[!htb]
\centering
\centerline{\includegraphics[width=8cm]{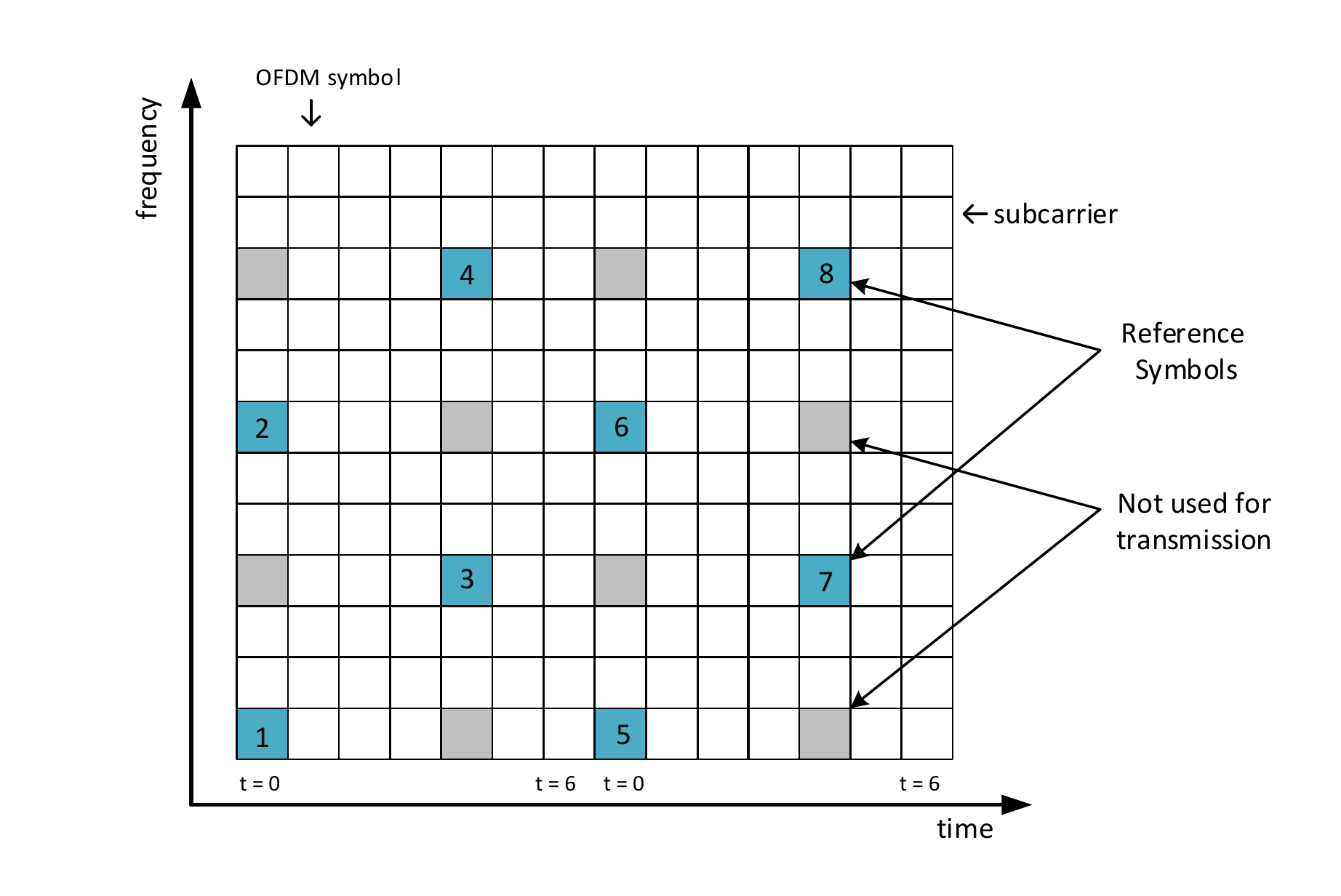}}
\vspace*{-5mm}
\caption{\small{LTE downlink (frequency domain) reference symbol layout of antenna port 0 in a two-antenna system.}}
\label{fig:Ref_LTE}
\end{figure}

In LTE downlink the reference symbols (or pilots) are orthogonal in both time and frequency as illustrated in Fig.~\ref{fig:Ref_LTE}. Thus, 2D MMSE channel estimation for single-input single-output (SISO) system is applicable to each transmit-receive antenna pair \cite{van1995channel}. Without loss of generality, we will present the MMSE channel estimation procedure for a SISO system, thus dropping the spatial indices $i$ and $j$.

\vspace*{-3mm}

For frequency-domain channel estimation in one physical resource block (PRB), we unroll the time-frequency grid along the frequency-axis first and then along the time-axis.
Thus, according to the orders indicated in Fig.~\ref{fig:Ref_LTE}, the transmit and receive reference symbols can be arranged in column vectors $\mathbf{x}_p = [ x_1 \, x_2 \, \ldots \, x_8]^T$ and $\mathbf{y}_p = [ y_1 \, y_2 \, \ldots \, y_8]^T $, respectively.
Let $\mathbf{h}_{f,\ell} \in \mathbb{C}^{N_c \times 1}$ denote the frequency-domain channel vector at time $\ell$, and $\mathbf{h}_f \in \mathbb{C}^{N_o N_c \times 1}$ denote unrolled channel vector in one subframe by stacking $\mathbf{h}_{f,\ell}$'s, 
i.e., $\mathbf{h}_{f} = [\mathbf{h}_{f,1}^T, \ldots, \mathbf{h}_{f,N_o}^T]^T$, where $N_o$ is the number of OFDM symbols in one subframe.
%\footnote{We caution that the system model in Section \ref{sec:sys_model} describes the transmission in one OFDM symbol period only. For example, the $\ell$-th OFDM symbol.}.
Also, let $\mathbf{h}_{t,\ell}$ and $\mathbf{h}_{t}$ be the time-domain counterparts corresponding to $\mathbf{h}_{f,\ell}$ and $\mathbf{h}_{f}$, respectively.
Dictated by the multiplicative nature as seen in Eq.~(\ref{eq:rx_sig2}), the CPE term $a[0]$ is lumped with channel coefficients to form an effective channel $\mathbf{h}'_{f,\ell} = a_{\ell}[0] \mathbf{h}_{f,\ell} $ and so is $\mathbf{h}'_{f}$. Moreover, the ICI term is treated as part of the noise process.
Therefore, the least squares (LS) channel estimates on pilot tones are $\widehat{ \mathbf{h}_{f,p}^{'LS} } = \mathbf{y}_p ./ \mathbf{x}_p$, where the MATLAB notation $./$ stands for element-wise division.

The MMSE estimator that uses channel second-order statistics is given by \cite{hou2010novel},
\begin{equation} \label{eq:ch_mmse}
\widehat{ \mathbf{h}_{f}^{'MMSE} } = \mathbf{R}_{h'h'_p} (\mathbf{R}_{h'_p h'_p} + \mu \sigma^2 \mathbf{I})^{-1} \widehat{ \mathbf{h}_{f,p}^{'LS} },
\end{equation}
where $\mathbf{R}_{h'_p h'_p}$ represents correlation matrix for channels on pilot tones and $\mathbf{R}_{h'h'_p}$ denotes the correlation between channels on all tones and pilot tones.
Here, the constant $\mu = \mathbb{E}\{ 1/|x_i[k]|^2 \}$ is constellation-dependent and the lumped noise variance $\sigma^2 = \sigma_w^2 + \sigma_{ICI}^2$ can be estimated from the null subcarriers in the system guard band \cite{wu2002phase}.
Note that, different from the 2-stage approach in \cite{hou2010novel}, the above correlation matrices include both frequency and time correlations. 
Specifically,
\begin{equation}
\mathbf{R}_{h' h'} \triangleq \mathbb\{  \mathbf{h}_f' (\mathbf{h}_f')^H \} = \mathbf{R}_F \otimes \mathbf{R}_T,
\end{equation}
where $\otimes$ denotes the Kronecker product, and $\mathbf{R}_F$ incorporates frequency correlation while $\mathbf{R}_T$ incorporates time correlation.
In the sequel, we will elaborate on how to obtain $\mathbf{R}_{h' h'}$ in such a form, and how to generate $\mathbf{R}_F $ and $\mathbf{R}_T$ by use of the statistics of $a[0]$.

Now consider the $(\ell,m)$ subblock of $\mathbf{R}_{h' h'}$, given by
\begin{equation} \label{eq:blk_corr}
\begin{split}
\mathbf{R}_{h' h'} (\ell,m)  = &  \; \mathbb{E} \left\{ a_{\ell}[0] \mathbf{h}_{f,\ell} (a_{m}[0] \mathbf{h}_{f,m})^H \right\} \\
=  & \; \mathbb{E} \left\{ a_{\ell}[0] a_m^*[0] \right\}   \mathbb{E} \left\{ \mathbf{h}_{f,\ell}  \mathbf{h}_{f,m}^H \right\}  \\
= & \; \mathbb{E} \left\{ a_{\ell}[0] a_m^*[0] \right\}  \mathbf{F}  \mathbb{E} \left\{\mathbf{h}_{t,\ell} \mathbf{h}_{t,m}^H \right\} \mathbf{F}^H,
\end{split}
\end{equation}
where $\mathbf{h}_{t,\ell} = [h_{t,\ell}[0], \ldots, h_{t,\ell}[L-1]]^T$ with $h_{t,\ell}[i]$ being the $i$-th channel tap at the $\ell$-th OFDM symbol,
and $\mathbf{F}$ is the DFT matrix of appropriate size.
Since we assume wide-sense stationary uncorrelated scattering channel, the channel taps at different delays are uncorrelated. Therefore,
\begin{equation} \label{eq:single_corr}
\mathbb{E}\{ h_{t,\ell}[i] h_{t,m}^*[j] \} = \left\{ 
\begin{array}{cl}
J_0 ( 2 \pi f_d |\ell - m| T_o ) \gamma_i,   & i = j \\
0,   & i \neq j
\end{array}\right.
\end{equation}
where $J_0(\cdot)$ is the zero-order Bessel function of the first kind, $f_d$ is the Doppler frequency, $T_o$ is the time duration of one OFDM symbol 
and $\gamma_i = \mathbb{E}\{ |h_{t,\ell}[i]|^2 \}$ is the power of the $i$-th channel tap given by power-delay profile (PDP). 
We note that PDP is statistical information that can be measured offline, and it is assumed to be known.

By substituting Eq.~(\ref{eq:single_corr}) into Eq.~(\ref{eq:blk_corr}), we can get
\begin{equation}
\mathbf{R}_{h' h'} (\ell,m) = J_0 ( 2 \pi f_d |\ell - m| T_o )   \mathbb{E} \left\{ a_{\ell}[0] a_m^*[0] \right\}  \mathbf{F} \Gamma  \mathbf{F}^H
\end{equation}
where $\Gamma = \text{diag}(\gamma_0, \ldots, \gamma_{L-1})$ is a diagonal matrix. 
We recognize that the frequency auto-correlation matrix $\mathbf{R}_F$ is given by
\begin{equation} \label{eq:freq_corr}
\mathbf{R}_F  =  \mathbb{E} \left\{ \mathbf{h}_{f,\ell}  \mathbf{h}_{f,\ell}^H \right\}  
= \mathbf{F}  \mathbb{E} \left\{\mathbf{h}_{t,\ell} \mathbf{h}_{t,\ell}^H \right\} \mathbf{F}^H = \mathbf{F} \Gamma  \mathbf{F}^H
\end{equation}
and we define the $(\ell,m)$ entry of the temporal auto-correlation matrix $\mathbf{R}_T$ by
\begin{equation}
\mathbf{R}_{T} (\ell,m) = J_0 ( 2 \pi f_d |\ell - m| T_o )   \mathbb{E} \left\{ a_{\ell}[0] a_m^*[0] \right\}
\end{equation}
where the computation of $\mathbb{E} \left\{ a_{\ell}[0] a_m^*[0] \right\}$ is shown in Appendix \ref{app:a0_corr}.
When $\ell = m$, $\mathbb{E} \left\{ |a_{\ell}[0]|^2 \right\} \approx 1 - 2 \pi \beta T_s N_t / 3$ \cite{schenk2008rf},
and for $\ell \neq m$, 
\begin{equation} \label{eq:a0_corr}
\mathbb{E} \left\{ a_{\ell}[0] a_m^*[0] \right\}  =
\frac{e^{-2 \pi \beta T_s N_c \vert m - \ell \vert}}{N_c^2} \cdot \frac{1- \cos(2\pi\beta T_s N_c)}{1- \cos(2\pi\beta T_s)}.
\end{equation}
Finally, $\mathbf{R}_{h' h'} (\ell,m) = \mathbf{R}_{T} (\ell,m) \mathbf{R}_F$ and thus $\mathbf{R}_{h' h'} = \mathbf{R}_{T} \otimes \mathbf{R}_F$.
Then, the correlation matrices $\mathbf{R}_{h' h'_p}$ and $\mathbf{R}_{h'_p h'_p}$ are generated by selecting corresponding time-frequency slots from $\mathbf{R}_{h' h'}$.

\section{Iterative Detection and Estimation Algorithm}

%1. Objective function \\
In this section, we will present a PN mitigation algorithm that iteratively updates the estimates of channel, phase noise and data symbols. The primary objective of interest is to minimize the squared-error $\Vert \mathbf{y} - (\mathbf{A} \otimes \mathbf{I}_{N_r}) \mathbf{H} \mathbf{x} \Vert^2$.
Due to the unknown phase noise matrix $\mathbf{A}$ and channel matrix $\mathbf{H}$,  ${\mathbf{x}}$ cannot be recovered directly via this objective function. Instead, we modify the objective function as follows
\begin{equation} \label{eq:obj}
\Vert \mathbf{y} - \left( \mathbf{A} \otimes \mathbf{I}_{N_r} \right) \mathbf{H} \mathbf{x} \Vert^2 =
\Vert \mathbf{y} - \left( \frac{\mathbf{A}}{a[0]} \otimes \mathbf{I}_{N_r} \right) \, a[0]\mathbf{H} \mathbf{x} \Vert^2,
\end{equation}
where $\mathbf{H}' = a[0]\mathbf{H}$ is available through the 2D MMSE channel estimation in Section \ref{sec:2d_mmse}, and the modified phase noise matrix $\mathbf{A}' = \mathbf{A}/{a[0]}$ is initialized to $\mathbf{I}_{N_c}$, based on which we can perform CPE compensation.
%With the initialized $a[0]\mathbf{H}$ and $\mathbf{A}/{a[0]}$, we can start MIMO detection by solving the least squares (LS) problem in Eq.~(\ref{eq:obj}).

However, the initial estimate of $\hat{\mathbf{x}}$ is fairly coarse, and the elements on the diagonal band of matrix $\mathbf{A}$ are not negligible, though $a[0]$ is dominating.
Also, the effective channel $a[0] \mathbf{H}$ is estimated in presence of high-level noise (lumped ICI and channel noise).
In consideration of these facts, we propose to iteratively update the three quantities, namely, phase noise matrix $\mathbf{A}'$, channel matrix $\mathbf{H}'$ and signal vector $\mathbf{x}$, one after another following the LS principle.
%We use the same LS principle to update all the quantities one by one -- MIMO detection of $\mathbf{x}$ based on initialized $\hat{\mathbf{H}}'$ and $\hat{\mathbf{A}}'$, followed by phase noise estimation using estimated $\hat{\mathbf{H}}'$ and $\hat{\mathbf{x}}$, and then update $\mathbf{H}'$ by use of latest estimation $\hat{\mathbf{A}}'$ and $\hat{\mathbf{x}}$.

%2. Matrix manipulations \\
For evaluating frequency-domain phase noise, we rewrite the transmit-receive relation in Eq.~(\ref{eq:rx_sig1}) as follows
\begin{equation}
\mathbf{y}  = (\mathbf{A}' \otimes \mathbf{I}_{N_r}) \mathbf{H}' \mathbf{x} + \mathbf{w}
 = \text{cir}(\mathbf{H}' \mathbf{x})_{N_r} \mathbf{a}' + \mathbf{w},
\end{equation}
where $\text{cir}(\mathbf{H}' \mathbf{x})_{N_r}$ means a cyclic shift of the vector $\mathbf{H}' \mathbf{x}$ by $N_r$ elements vertically, and the resulting matrix is of size $N_c N_r \times N_c$.
Here, we use the fact that $\mathbf{A}' = \text{cir}(\mathbf{a}')$, where $\mathbf{a}' = \mathbf{a}/a[0]$. Now, we obtain an over-determined system of $N_r N_c$ equations to estimate parameters $\mathbf{a}'$ of dimension $N_c$. For better estimation, we adopt the method in \cite{zou2007compensation} to further reduce the number of unknown parameters. Concretely, we estimate time-domain decimated phase noise samples $\mathbf{c}'$ instead of $\mathbf{a}'$, and they are related by
\begin{equation} \label{eq:pn_est}
\mathbf{a}' \approx \frac{1}{N_c} \mathbf{F}_a \mathbf{P} \mathbf{c}',
\end{equation}
where $\mathbf{F}_a$ is an $N_c \times N_c$ DFT matrix and $\mathbf{P}$ is a linear interpolation matrix of size $N_c \times M (M \ll N_c)$ as specified by Eq.~(15) in \cite{zou2007compensation}.

Next, the transmit-receive equation in Eq.~(\ref{eq:rx_sig1}) is rearranged to apply LS channel estimation
\begin{equation} \label{eq:ch_est}
\mathbf{y}  = (\mathbf{A}' \otimes \mathbf{I}_{N_r}) \mathbf{H}' \mathbf{x} + \mathbf{w}
= (\mathbf{A}' \otimes \mathbf{I}_{N_r}) \mathbf{X} \mathbf{h}' + \mathbf{w},
\end{equation}
where $\mathbf{h}' = [ (\mathbf{h}'_{11})^T \, (\mathbf{h}'_{12})^T \, \ldots \, (\mathbf{h}'_{ji})^T \ldots \, (\mathbf{h}'_{N_r N_t})^T ]^T$ and $\mathbf{h}'_{ji} = [H_{ji}[0] \, H_{ji}[1] \, \ldots \, H_{ji}[N_c-1] ]^T/a[0]$. In Eq.~(\ref{eq:ch_est}), the $N_c N_r$-by-$N_c N_t N_r$ matrix $\mathbf{X}$, consisting of elements from vector $\mathbf{x}$, can be concatenated as $\mathbf{X} = [\mathbf{X}_1\, \mathbf{X}_2 \, \ldots \mathbf{X}_j \ldots \, \mathbf{X}_{N_r}]$, where the submatrix $\mathbf{X}_j$ is specified as
\begin{equation}
\mathbf{X}_j(m,n) =  \left\{ \begin{array}{rcl}
       x_i[k],   & \mbox{if} &  m=k N_r + j, \\
                   &    & n=(i-1)N_c + k +1, \\
      0, &  & \mbox{otherwise}.
                \end{array}\right.
\end{equation}
Nevertheless, because $N_c N_r < N_c N_t N_r$ when $N_t > 1$, Eq.~(\ref{eq:ch_est}) might be an under-determined system. To overcome this difficulty, we reduce the number of unknowns by estimating channel coefficients in time domain instead of frequency-domain, since channel delay spread $L \ll N_c$ in practice. Extract the first $L$ columns of the DFT matrix $\mathbf{F}_a$ to form matrix $\mathbf{F}_h$, and denote $\mathbf{F}_H = \text{blkdiag}\{\mathbf{F}_h\}$ with $N_r N_t$ identical blocks $\mathbf{F}_h$. Thus, the time-domain channel coefficients $\mathbf{h}'_t$ is associated with $\mathbf{h}'$ by the relation $\mathbf{h}' = \mathbf{F}_H \mathbf{h}'_t$.

%3. Algorithm \\
\begin{algorithm}
\caption*{\textbf{Algorithm} Iterative Detection and Estimation (IDE)} \label{alg:IDE}
\begin{algorithmic}[1]
\State Obtain initial $\hat{\mathbf{H}}'$ by 2D MMSE channel estimation, and initialize $\hat{\mathbf{A}}' = \mathbf{I}_{N_c}$.
\Repeat
\State MIMO detection by LS estimation
\begin{equation} \label{eq:sym_det}
\hat{\mathbf{x}} = \operatorname*{arg\,min}_{\mathbf{x}} \left\Vert \mathbf{y} - (\hat{\mathbf{A}}' \otimes \mathbf{I}_{N_r}) \hat{\mathbf{H}'} \mathbf{x} \right\Vert^2,
\end{equation}
and then slice $\hat{\mathbf{x}}$ to the nearest constellation point.
\State Time-domain phase noise estimation is given by
\begin{equation}  \label{eq:pn_est}
\hat{\mathbf{c}}' = \operatorname*{arg\,min}_{\mathbf{c}'}
\left\Vert \mathbf{y} - \text{cir}(\hat{\mathbf{H}}' \hat{\mathbf{x}})_{N_r} \cdot \frac{1}{N_c} \mathbf{F}_a \mathbf{P} \mathbf{c}'  \right\Vert^2.
\end{equation}
Transform $\hat{\mathbf{c}}'$  back to frequency domain $\hat{\mathbf{a}}' = \mathbf{F}_a \mathbf{P} \hat{\mathbf{c}}' / N_c$, and normalize $\hat{\mathbf{a}}' = \hat{\mathbf{a}}' / \hat{a}[0]$. Then obtain
$\hat{\mathbf{A}}' = \text{cir} (\hat{\mathbf{a}}') $.
\State Update channel estimation in time domain
\begin{equation}
\hat{\mathbf{h}}'_t = \operatorname*{arg\,min}_{\mathbf{h}'_t}
\left\Vert \mathbf{y} - (\hat{\mathbf{A}}' \otimes \mathbf{I}_{N_r}) \hat{\mathbf{X}} \mathbf{F}_H \mathbf{h}'_t \right\Vert^2
\end{equation}
and then $\hat{\mathbf{h}}' = \mathbf{F}_H \hat{\mathbf{h}}'_t$. Finally, reshape $\hat{\mathbf{h}}'$ to $\hat{\mathbf{H}}'$.
\Until{Reach maximum iterations or no significant improvement of the objective function}.
\end{algorithmic}
\end{algorithm}

%4. LS complexity  \\
The above procedures are summarized in the iterative detection and estimation (IDE) algorithm.
We notice that each update step involves large-scale matrix multiplications and LS computations.
In practical LTE systems, $N_r, N_t \ll N_c$, and also $M, L \ll N_c$ by algorithm construction.
By fixing $N_r, N_t, M$ and $L$ as constant parameters, the complexity of the matrix multiplication and the LS computation using QR factorization \cite{golub2012matrix} are both $O(N_c^3)$ per iteration.
Therefore, the overall complexity is $O(N_c^3)$ with finite maximum iterations.
%We notice that the scale of the LS problems in above algorithm might be quite large, if the MIMO-OFDM system is equipped with plenty of transmit/receive antennas and/or with numerous subcarriers. Nonetheless, large-scale LS has been extensively studied in the matrix computation community \cite{golub2012matrix}. Two common approaches include the QR factorization and the Cholesky factorization of normal equations. It is generally concluded that the former method has better numerical properties while the latter one is faster.

\section{Numerical Results}
This section provides simulation results under LTE structure with $2\times2$ spatially independent MIMO in a 3 MHz channel bandwidth,
in which 180 out of 256 subcarriers are occupied and the rest are guard tones.
16-QAM is used for modulating the data symbols, and reference symbols are generated from the 4 outermost corner of the 16-QAM constellation.
The tone spacing is 15 KHz, and thus the sampling period $T_s = 1/(15000 \times 256)$.
Further, one subframe in LTE system, containing 14 OFDM symbols, lasts for 1 milli-second, thus the OFDM symbol period $T_o = 1/14$ ms.
The multipath fading channel used in the simulation is the ITU Pedestrian B (PedB) model, in which the tap delays are at $[0.0  \; 0.2 \; 0.8 \;  1.2 \;  2.3 \;  3.7]$ micro-second with power delay profile $[0.0 \; -0.9 \; -4.9 \; -8.0 \; -7.8 \; -23.9]$ measured in dB.
We assume pedestrian velocity at $v = 5$ km/h (= 5/3.6 m/s) and a carrier frequency $f_c = 2$ GHz.
Moreover, we set the number of decimated time-domain phase noise elements $M = 50$, and the maximum number of iterations for IDE algorithm is 5.

The proposed IDE scheme is compared with the algorithm presented for an LTE-like system in \cite{corvaja2009joint}.
However, the algorithm in \cite{corvaja2009joint} requires pilot symbols in every OFDM symbol, 
which is not the case for real LTE subframe structure. Thus, besides the frequency interpolation that the authors proposed, 
we perform another interpolation along the time axis to obtain CSI for OFDM symbols without training signals. 
Moreover, we compare the IDE algorithm with CPE compensation, which is described in Eq.~(\ref{eq:sym_det}) (essentially the first iteration of IDE algorithm) based on the MMSE channel estimates of Eq.~(\ref{eq:ch_mmse}).
Particularly, two channel estimates are considered for CPE compensation -- with and without the statistics of $a[0]$ in MMSE channel estimation.
Last but not least, two benchmarks, the cases of no compensation and no PN, are also plotted for comparison purpose.

\begin{figure}[!tb]
\centering
\centerline{\includegraphics[width=8cm]{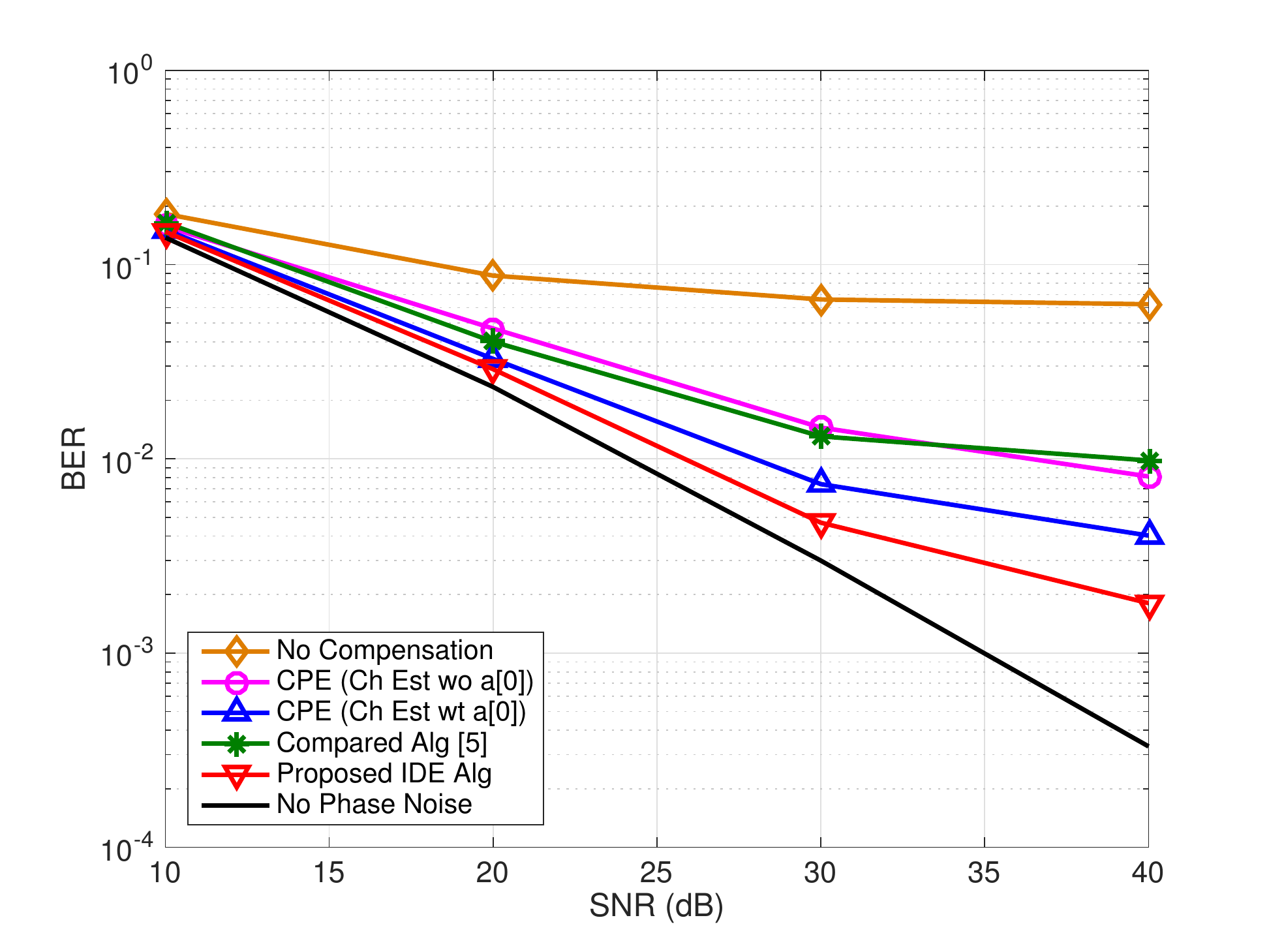}}
%\vspace*{-3mm}
\caption{\small{BER versus SNR at $\beta=25$Hz.}}
\label{fig:BER_SNR}
\end{figure}

\begin{figure}[!tb]
\centering
\centerline{\includegraphics[width=8cm]{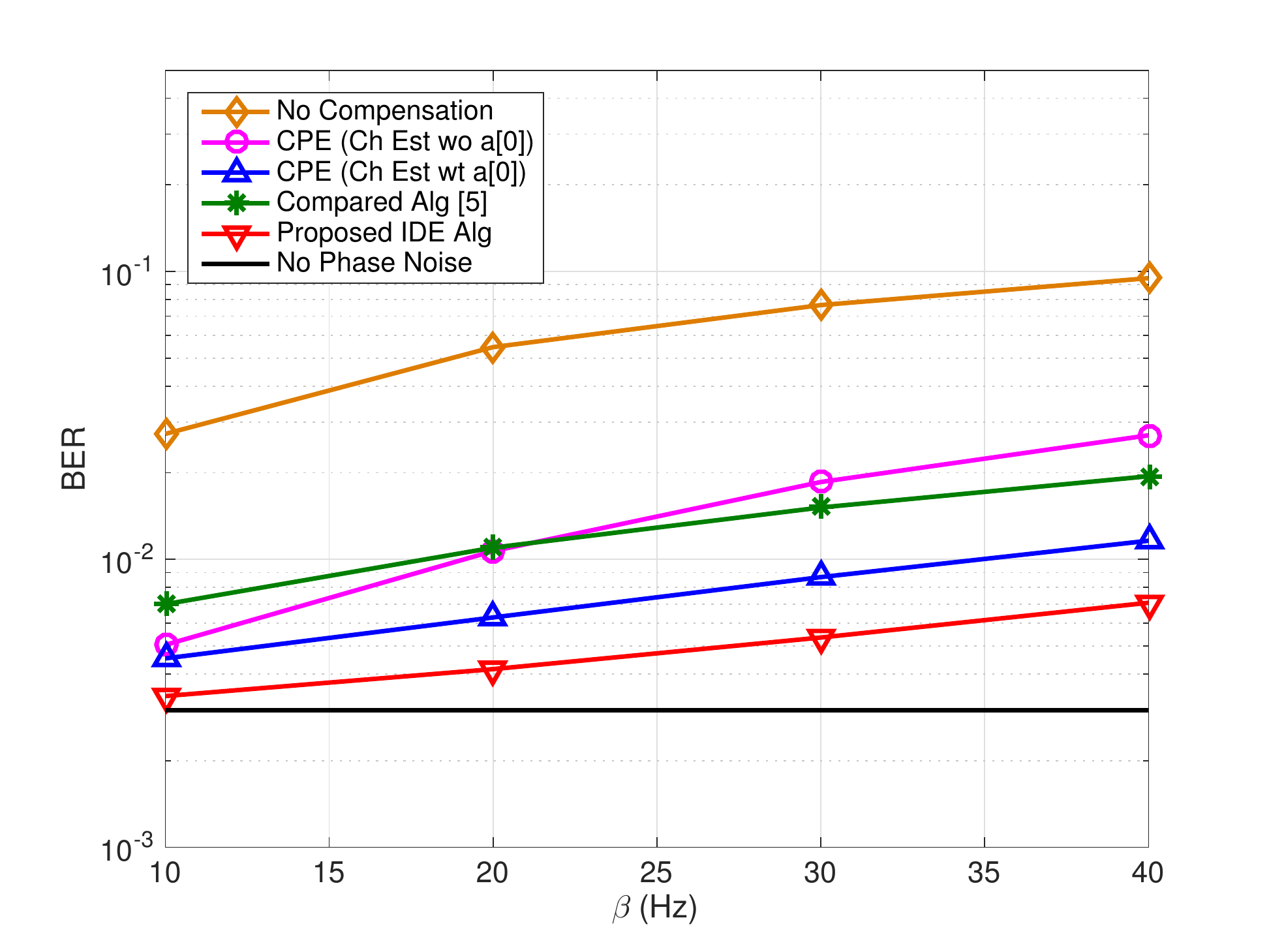}}
%\vspace*{-3mm}
\caption{\small{BER versus $\beta$ at SNR\;=\;30dB.}}
\label{fig:BER_Beta}
\end{figure}

\begin{figure}[!tb]
\centering
\centerline{\includegraphics[width=8cm]{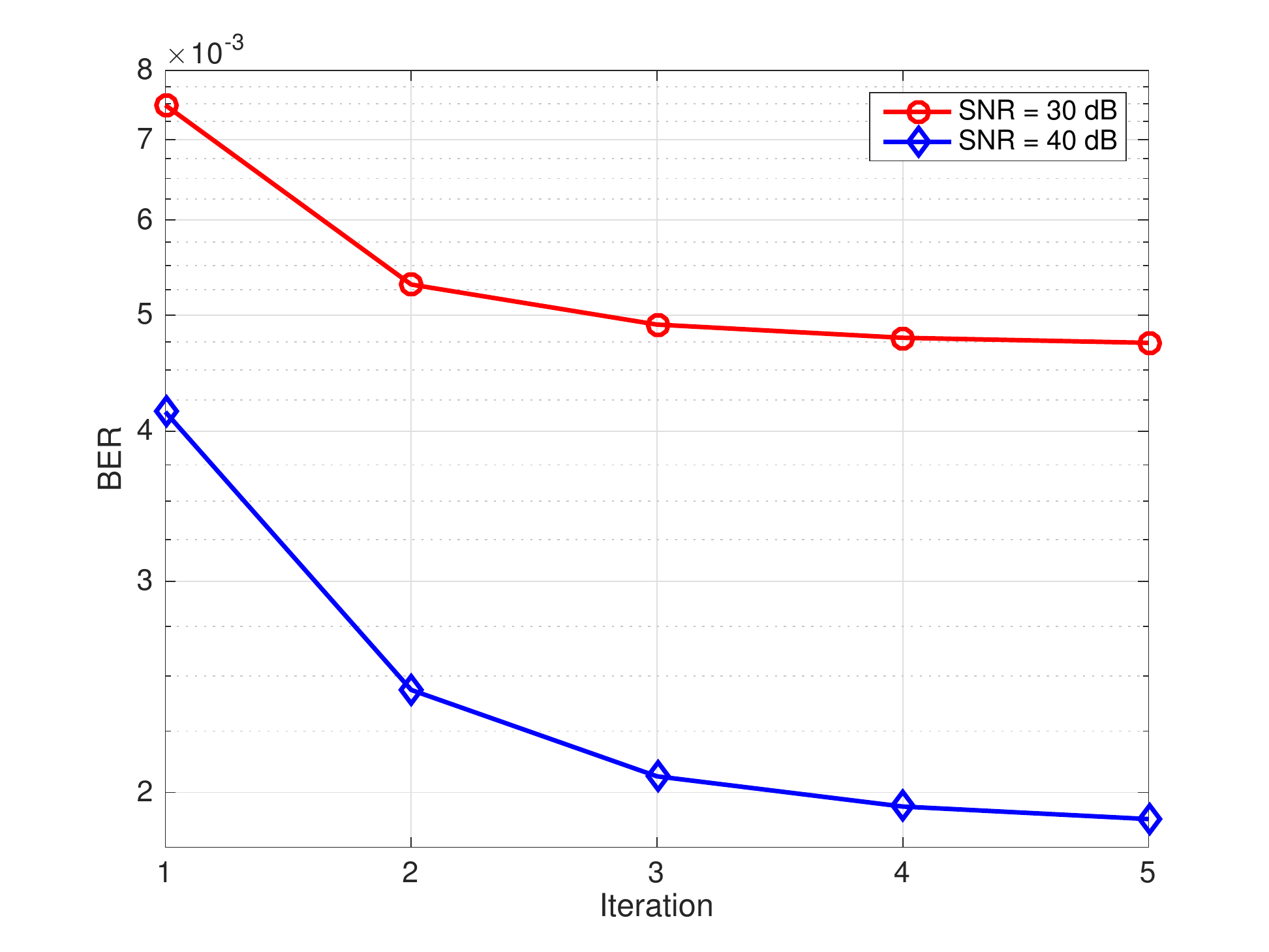}}
%\vspace*{-3mm}
\caption{\small{BER vs. number of iterations at $\beta=25$Hz.}}
\label{fig:BER_Iter}
\end{figure}

Fig.~\ref{fig:BER_SNR} shows the bit error rate (BER) performance against signal-to-noise ratio (SNR), while Fig.~\ref{fig:BER_Beta} demonstrates BER versus $\beta$.
It is clearly seen in Fig.~\ref{fig:BER_SNR} and Fig.~\ref{fig:BER_Beta} that the IDE algorithm outperforms all other algorithms in between the two benchmarks, and it is also observed that the CPE compensation with channel estimates considering $a[0]$ is superior to that without taking $a[0]$ into account in the channel estimation.
Further, it is interesting to note that the compared algorithm in \cite{corvaja2009joint} performs close to the inferior CPE compensation in Fig.~\ref{fig:BER_SNR}, though it outperforms the latter in Fig.~\ref{fig:BER_Beta} when $\beta$ is large.
The poor performance of the algorithm in \cite{corvaja2009joint} is possibly due to the sparser reference symbols (based on LTE setup) than those considered by the authors of \cite{corvaja2009joint} and also caused by the less accurate channel estimates by interpolation.
Moreover, since our proposed IDE algorithm is iterative, we show its BER performance as iterations increase. 
It is clear from Fig.~\ref{fig:BER_Iter} that 5 iterations will suffice.

%\vspace*{-3mm}

\section{Conclusion}
In this correspondence, we proposed an iterative phase noise and channel estimation and compensation algorithm that is well suited to the wireless systems with sparse pilots arrangement.
This algorithm starts with a fairly accurate channel estimate, and then follows the LS principle to perform iterative updates.
Its effectiveness has been shown via computer simulations.
To combat imperfections, joint detection-decoding schemes have been proposed recently using LDPC codes \cite{wang2016diversity,wang2014joint,wang2015joint,wang2016robust,wang2015diversity,wang2016fec}.
Specially, the iterative methods in \cite{wang2017galois,wang2018joint} may be used for more effective phase noise compensation. 
Moreover, the proposed IDE algorithm is heuristic in nature, 
so future works may consider a better termination criterion with provable performance bounds.

% if have a single appendix:
%\appendix[Proof of the Zonklar Equations]
% or
%\appendix  % for no appendix heading
% do not use \section anymore after \appendix, only \section*
% is possibly needed

% use appendices with more than one appendix
% then use \section to start each appendix
% you must declare a \section before using any
% \subsection or using \label (\appendices by itself
% starts a section numbered zero.)
%

\appendices
\section{Correlation of  $a_{\ell}[0]$ and $a_m[0]$} \label{app:a0_corr}

The carrier process autocorrelation follows \cite{schenk2008rf}
\begin{equation}
R_{aa} = \mathbb{E} \left[ e^{j \phi[n+k]} (e^{j \phi[n]})^*  \right]
= \exp(-2 \pi \beta T_s |k| ).
\end{equation}
Given $R_{aa}$, the correlation $\mathbb{E} \left\{ a_{\ell}[0] a_m^*[0] \right\} $ for $m > \ell$ is

\begin{equation}
\begin{split}
& \; \mathbb{E} \left\{ a_{\ell}[0] a_m^*[0] \right\}  \\
%= & \; \mathbb{E} \left\{ \frac{1}{N_c} \sum_{p=N_c\ell}^{N_c\ell+N_c-1} e^{j \phi[p]} \cdot \frac{1}{N_c} \sum_{q = N_cm}^{N_cm+N_c-1} e^{-j \phi[q]} \right\}  \\
= & \; \frac{1}{N_c^2} \sum_{p=N_c\ell}^{N_c\ell+N_c-1} \sum_{q = N_cm}^{N_cm+N_c-1} \mathbb{E} \left\{ e^{j \phi[p]}  e^{-j \phi[q]} \right\} \\
= & \; \frac{1}{N_c^2} \sum_{p=0}^{N_c-1} \sum_{q=0}^{N_c-1} \mathbb{E} \left\{ e^{j \phi[N_c\ell+p]} e^{-j \phi[N_cm+q]}  \right\} \\
= & \; \frac{1}{N_c^2} \sum_{p=0}^{N_c-1} \sum_{q=0}^{N_c-1} e^{-2 \pi \beta T_s \vert N_cm+q - N_c\ell - p \vert }  \\
%= & \; \frac{ e^{-2 \pi \beta T_s N_c (m - \ell)} }{N_c^2} \cdot
%\sum_{p=0}^{N_c-1} \sum_{q=0}^{N_c-1} e^{-2 \pi \beta T_s ( q - p)} \\
= & \; \frac{ e^{-2 \pi \beta T_s N_c (m - \ell)} }{N_c^2} \cdot
\sum_{p=0}^{N_c-1} e^{2 \pi \beta T_s p} \cdot \sum_{q=0}^{N_c-1} e^{-2 \pi \beta T_s q} \\
%= & \; \frac{e^{-2 \pi \beta T_s N_c (m - \ell)}}{N_c^2} \cdot \frac{1- e^{2\pi\beta T_s N_c}}{1- e^{2\pi\beta T_s}} \cdot \frac{1- e^{- 2\pi\beta T_s N_c}}{1- e^{- 2\pi\beta T_s}}  \\
= & \; \frac{e^{-2 \pi \beta T_s N_c (m - \ell)}}{N_c^2} \cdot \frac{1- \cos(2\pi\beta T_s N_c)}{1- \cos(2\pi\beta T_s)}.
\end{split}
\end{equation}
Considering the symmetric case for $m < \ell$, we obtain
\begin{equation*}
\mathbb{E} \left\{ a_{\ell}[0] a_m^*[0] \right\} =
\frac{e^{-2 \pi \beta T_s N_c \vert m - \ell \vert}}{N_c^2} \cdot \frac{1- \cos(2\pi\beta T_s N_c)}{1- \cos(2\pi\beta T_s)},
\end{equation*}
for any $\ell \neq m$.

% ============ Simulation Figures ==============
%\begin{figure}[!htb]
%\centering
%\centerline{\includegraphics[width=8cm]{Figs/Fig1_BER_vs_SNR.eps}}
%\caption{BER Comparisons versus SNR using 16-QAM and $\beta=40$Hz.}
%\label{fig:BER_SNR}
%\end{figure}

%\begin{figure}[!htb]
%\centering
%\centerline{\includegraphics[width=8cm]{Figs/Fig2_MSE_vs_SNR.eps}}
%\caption{MSE Comparisons versus SNR using 16-QAM and $\beta=40$Hz.}
%\label{fig:MSE_SNR}
%\end{figure}

%\begin{figure}[!htb]
%\centering
%\centerline{\includegraphics[width=10cm]{Figs/Fig3_BER_vs_Beta.eps}}
%\caption{BER Comparisons versus $\beta$ using 16-QAM and SNR = 40dB.}
%\label{fig:BER_Beta}
%\end{figure}
% ========================================

% use section* for acknowledgement
%\section*{Acknowledgment}
%
%
%The authors would like to thank...

% Can use something like this to put references on a page
% by themselves when using endfloat and the captionsoff option.
\ifCLASSOPTIONcaptionsoff
  \newpage
\fi

% trigger a \newpage just before the given reference
% number - used to balance the columns on the last page
% adjust value as needed - may need to be readjusted if
% the document is modified later
%\IEEEtriggeratref{8}
% The "triggered" command can be changed if desired:
%\IEEEtriggercmd{\enlargethispage{-5in}}

% references section

% can use a bibliography generated by BibTeX as a .bbl file
% BibTeX documentation can be easily obtained at:
% http://www.ctan.org/tex-archive/biblio/bibtex/contrib/doc/
% The IEEEtran BibTeX style support page is at:
% http://www.michaelshell.org/tex/ieeetran/bibtex/
%
%\section{REFERENCES}
%\label{sec:refs}

\bibliographystyle{IEEEtran}
\bibliography{IEEEabrv,myPNbibfile}

%
% <OR> manually copy in the resultant .bbl file
% set second argument of \begin to the number of references
% (used to reserve space for the reference number labels box)
%\begin{thebibliography}{1}
%
%\bibitem{IEEEhowto:kopka}
%H.~Kopka and P.~W. Daly, \emph{A Guide to \LaTeX}, 3rd~ed.\hskip 1em plus
%  0.5em minus 0.4em\relax Harlow, England: Addison-Wesley, 1999.
%
%\end{thebibliography}

% that's all folks
\end{document}